\documentclass[runningheads]{article}
\usepackage{makeidx}  
\usepackage{amsmath}
\usepackage{amssymb,float}
\usepackage {hyperref} 
\usepackage[letterpaper,hmargin=1.22in,vmargin=1.1in]{geometry}
\def\ID-MQV{\textsf{ID-MQV}}
\def\eMB{\textsf{eMB}}
\def\nID-SYL{\textsf{nID-SYL}}
\begin{document}
%
%
%
%
%
\title{On the Relations Between Diffie-Hellman and  ID-Based Key Agreement from Pairings
\thanks{First version, January 2008; This version (July 2009) is a
minor revison.}}
\date{}
\author{Shengbao Wang \thanks{(Email: shengbaowang@gmail.com) The author is currently with New Star Institute of Applied Technology, China
}\\\\
}%

\maketitle              

\begin{abstract}
This paper studies the relationships between the traditional
Diffie-Hellman key agreement protocol and the identity-based
(ID-based) key agreement protocol from pairings.

For the Sakai-Ohgishi-Kasahara (SOK) ID-based key construction, we
show that identical to the Diffie-Hellman protocol, the SOK key
agreement protocol also has three variants, namely \emph{ephemeral},
\emph{semi-static} and \emph{static} versions. Upon this, we build
solid relations between authenticated Diffie-Hellman (Auth-DH)
protocols and ID-based authenticated key agreement (IB-AK)
protocols, whereby we present two \emph{substitution rules} for this
two types of protocols. The rules enable a conversion between the
two types of protocols. In particular, we obtain the \emph{real}
ID-based version of the well-known MQV (and HMQV) protocol.

Similarly, for the Sakai-Kasahara (SK) key construction, we show
that the key transport protocol underlining the SK ID-based
encryption scheme (which we call the ``SK protocol") has its non-ID
counterpart, namely the Hughes protocol. Based on this observation,
we establish relations between corresponding ID-based and
non-ID-based protocols.
In particular, we propose a highly enhanced version of the McCullagh-Barreto protocol.\\


\end{abstract}

%
\section{Introduction}
In 2005, Boyd and Choo \cite{BC05} and Wang \emph{et al.}
\cite{WCB05} noticed that there are some similarities between
(pairing-based) ID-based and non-ID-based authenticated key
agreement (AK) protocols. This study further investigate this
observation. Interestingly, we discover much more than those
researchers previously might imagined.

\subsection{Proposed Novel Protocols}
We discover some important \emph{substitution rules} (see Table
\ref{Tab:SOK}, \ref{Tab:SK}) between the two different types of
protocols. The rules enable a useful conversion between the
authenticated version of the two types of protocols. By applying
these rules, we present three novel protocols (namely, the protocols
which are highlighted in bold in Table \ref{Tab:2} and \ref{Tab:3})
which possesses remarkable performance and security.

\begin{figure}[H]
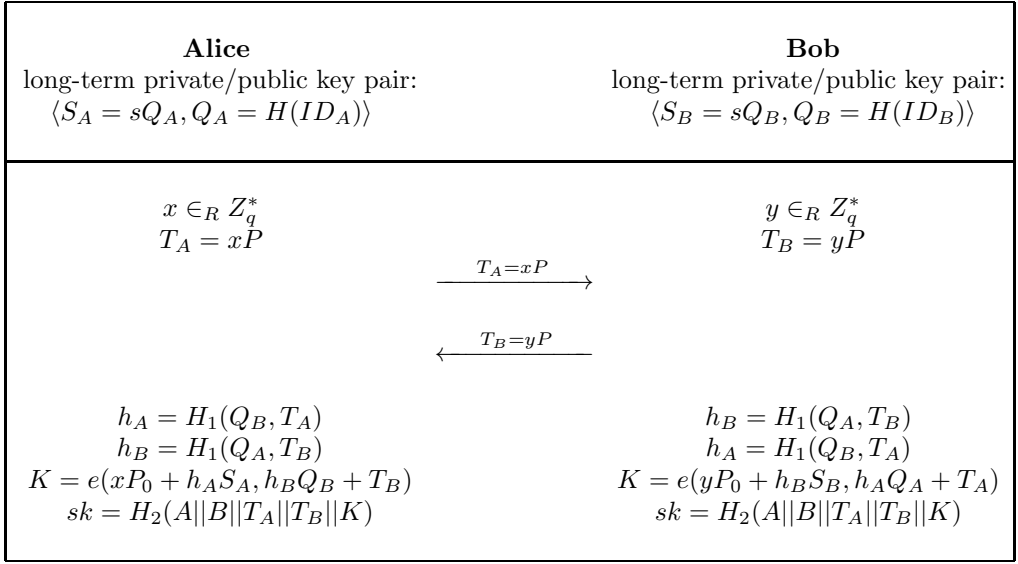

\center
\renewcommand{\arraystretch}{1}
\setlength\tabcolsep{0pt}

\begin{tabular}{|ccc|}
\hline
&&\\
\textbf{ Alice} & & \ \ \textbf{Bob}  \\
\ \ long-term private/public key pair:    & &   \ \ long-term private/public key pair:\ \ \\
 $\langle S_A = sQ_A, Q_A=H(ID_A)\rangle$   & &   \ \ $\langle S_B = sQ_B, Q_B=H(ID_B)\rangle$\\
 &&\\
\hline
 &&\\
 $x \in_R Z_q^\ast$  & &  \ \ $y \in_R Z_q^\ast$\\

$T_A = xP$   & &   \ \ $T_B = yP$\\

                 &  \ \ $\xrightarrow{\ \ \ \ T_{A}=xP\ \ \ \ }$ & \\
                 &&\\

                &  \ \ $\xleftarrow{\ \ \ \ T_{B}=yP\ \ \ \ }$  & \\
                 & & \\

\ \ $h_A = H_1(Q_B, T_A)$ & &  \ \ $h_B = H_1(Q_A, T_B)$\ \  \\
\ \ $h_B = H_1(Q_A, T_B)$ & &  \ \ $h_A = H_1(Q_B, T_A)$\ \  \\
 \ \ $K = e(x P_{0} + h_A S_A, h_B Q_B + T_B)$ & &  \ \ $K = e(y P_{0} + h_BS_B, h_AQ_A + T_A)$\ \  \\
 \ \ $sk = H_2(A||B||T_A||T_B||K)$ & &  \ \ $sk = H_2(A||B||T_A||T_B||K)$\ \  \\

&&\\
\hline
\end{tabular}\\

\caption{\ID-MQV: ID-Based (H)MQV Protocol}\label{Fig:ID-MQV}
\end{figure}

\begin{enumerate}
  \item The real ID-based version of the MQV (and, HMQV) protocol ---
  \ID-MQV. (See Fig. \ref{Fig:ID-MQV}.)
  \item The enhanced MB (McCullagh--Barreto) ID-based protocol ---
  \eMB. (See Fig. \ref{Fig:Enhanced-MB}.)
  \item The non-ID-based version of the SYL protocol ---
  \nID-SYL (See Appendix \ref{Appendix:A}, Fig. \ref{Fig:nonid-SYL} ).
 \end{enumerate}

\begin{table}[H]%
  \centering
  \caption{Corresponding Protocols (non-ID-Based \emph{vs.} ID-Based)}\label{Tab:2}
  \begin{tabular}{|c||c|ccc|}
    \hline
    Protocol Type & Prot. Message & Auth. DH Protocols & $\Leftrightarrow$ & ID-Based Protocols\\
    \hline\hline
    &&&&\\
         A0       & $T_A=xP$                 & MTI/A0             & $\Leftrightarrow$ &  Smart \cite{Sm02}            \\
   Enhanced A0    &                          & (H)MQV             & $\Leftrightarrow$ &  \textbf{\ID-MQV} (See Fig. \ref{Fig:ID-MQV})            \\
   \hline
   &&&&\\
         A1       & $T_A=xQ_A$               & MTI/A1             & $\Leftrightarrow$ &  Chen--Kudla \cite{CK02}     \\
   Enhanced A1    &                          & (H)MQV-1           & $\Leftrightarrow$ &  Wang \cite{Wangyongge05}, Chow--Choo \cite{CC07}            \\
   \hline
   &&&&\\
         C0       & $T_A=xQ_B$               & MTI/C0             & $\Leftrightarrow$ &  MB-1 \cite{MB05a} \\
   Enhanced C0    &                          & ECKE-1N \cite{WCSW08}  & $\Leftrightarrow$ & \textbf{ \textsf{eMB}} (See Fig. \ref{Fig:Enhanced-MB})            \\
         B0       &                          & MTI/B0             & $\Leftrightarrow$ &  MB-2 \cite{MB05b} \\
   \hline
   &&&&\\
         C1       & $T_A=xF_{AB}$            & MTI/C1             & $\Leftrightarrow$ &  Scott \cite{Sc02}            \\
   Enhanced C1    &                          & Enhanced MTI/C1 (See Fig. \ref{Fig:Enhanced-C1})  & $\Leftrightarrow$ &  Open Problem! \\

    \hline
  \end{tabular}

\end{table}

\begin{table}[H]%
  \centering
  \caption{Corresponding Protocols (Broken and Repaired Ones)}\label{Tab:3}
  \begin{tabular}{|c||c|ccc|}
    \hline
    Protocol Type &\ \ \ Protocol Message \ \ \  & \ \ \ Auth. DH Protocols &\ \ \ $\Leftrightarrow$\ \ \  & ID-Based Protocols\ \ \ \\
    \hline\hline
    &&&&\\
         A0 Variant-1   & $T_A=xP$           & Reduced MQV        & $\Leftrightarrow$ &  Shim \cite{Sh03}            \\
   Repaired Protocol   &                    & \textbf{\nID-SYL} (See Fig. \ref{Fig:nonid-SYL})  & $\Leftrightarrow$ &  SYL \cite{YL05}            \\
   \hline
   &&&&\\
         C0 Variant-1  & $T_A=xQ_B$         & $K=(x+y+xy)P$        & $\Leftrightarrow$ &  Xie \cite{Xie05b} \\
   Repaired Protocol   &                    & $K=(x+y)P||xyP$       & $\Leftrightarrow$ &  LYL \cite{LYL05}            \\
    \hline
  \end{tabular}

\end{table}

\section{Preliminaries}

\subsection{Bilinear Pairings}
Let $\mathbb{G}_1$ denotes an additive group of prime order $q$ and
$\mathbb{G}_2$ a multiplicative group of the same order. We let $P$
denote a generator of $\mathbb{G}_1$. For us, an admissible pairing
is a map $e: \mathbb{G}_1\times \mathbb{G}_1 \to \mathbb{G}_2$ with
the following properties:
\begin{enumerate}
    \item The map $e$ is bilinear: given $Q,R \in {\mathbb{G}}_1 $ and
$a,b \in Z_q^\ast $, we have $e(aQ,bR) = e(Q,R)^{ab}$.
    \item The map $e$ is non-degenerate: $e(P,P)\neq
1_{\mathbb{G}_2}$.
    \item The map $e$ is efficiently computable.
\end{enumerate}

Typically, the map $e$ will be derived from either the Weil or Tate
pairing on an elliptic curve over a finite field.

\section{Three Versions of the SOK Protocol and the Substitution Rules}

We first focus on the SOK ID-based key setting \cite{SOK00}. We show
that the \emph{static} SOK protocol from \cite{SOK00} has two more
variants, \emph{i.e.}, the \emph{semi-static} and \emph{ephemeral}
SOK protocols.

Note that the figures given in the rest of the paper are all
self-explaining.

\subsection{Static DH and the SOK-NIKD Protocols}
As observed by Boyd, Mao and Paterson \cite{BMP04} and Ryu \emph{et
al.} \cite{RYY04}, the two \emph{non-interactively} shared static
secret from the Diffie-Hellman protocol \cite{DH76} and the SOK
non-interactive ID-based key distribution (SOK-NIKD) protocol
\cite{SOK00} are $F_{DH}=abP$ and $F_{SOK}=e(Q_A, Q_B)^s$,
respectively.

\begin{figure}[H]
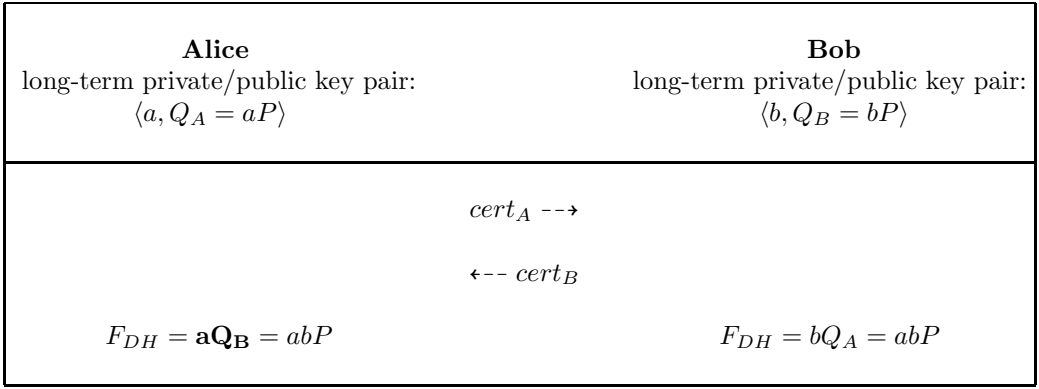

\center
\renewcommand{\arraystretch}{1}
\setlength\tabcolsep{0pt}

\begin{tabular}{|ccc|}
\hline
&&\\
\textbf{ Alice} & & \ \ \textbf{Bob}  \\
\ \ long-term private/public key pair:    & &   \ \ long-term private/public key pair:\ \ \\
 $\langle a, Q_A = aP \rangle$   & &   \ \ $\langle b, Q_B = bP\rangle$\\
 &&\\
\hline
 &&\\
                 &  \ \ $\ \ \ \ cert_{A}\dashrightarrow\ \ \ \ $ & \\
                 &&\\

                &  \ \ $\ \ \ \ \dashleftarrow{ cert_{B}\ \ \ \ }$  & \\
                 & & \\

\ \ $F_{DH} = \mathbf{aQ_B}=abP$ & &  \ \ $F_{DH}  = bQ_A = abP$\ \  \\

&&\\
\hline

\end{tabular}
\caption{The Static DH Protocol \cite{DH76}}\label{Fig:Static DH}
\end{figure}


\begin{figure}[H]
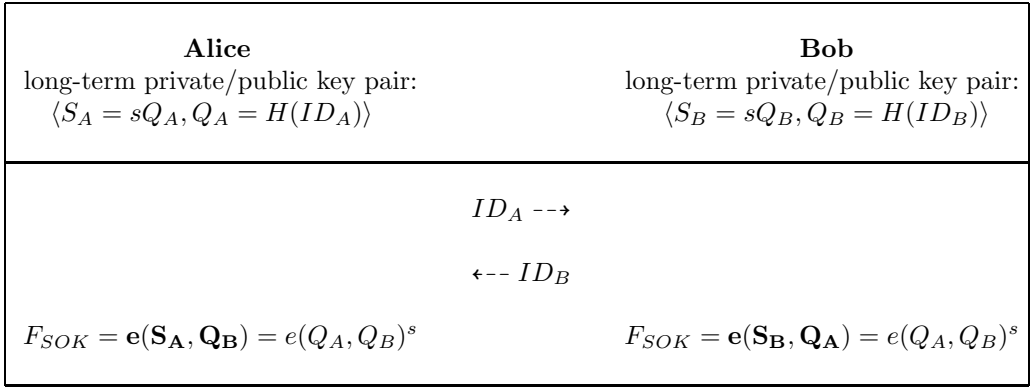

\center
\renewcommand{\arraystretch}{1}
\setlength\tabcolsep{0pt}

\begin{tabular}{|ccc|}
\hline
&&\\
\textbf{ Alice} & & \ \ \textbf{Bob}  \\
\ \ long-term private/public key pair:    & &   \ \ long-term private/public key pair:\ \ \\
 $\langle S_A = sQ_A, Q_A=H(ID_A)\rangle$   & &   \ \ $\langle S_B = sQ_B, Q_B=H(ID_B)\rangle$\\
 &&\\
\hline
 &&\\
                 &  \ \ $\ \ \ \ ID_{A}\dashrightarrow\ \ \ \ $ & \\
                 &&\\

                &  \ \ $\ \ \ \ \dashleftarrow{ ID_{B}\ \ \ \ }$  & \\
                 & & \\

\ \ $F_{SOK} = \mathbf{e(S_A, Q_B)}=e(Q_A, Q_B)^s$ & &  \ \ $F_{SOK} = \mathbf{e(S_B, Q_A)}=e(Q_A, Q_B)^s$\ \  \\

&&\\
\hline

\end{tabular}
\caption{The SOK-NIKD Protocol \cite{SOK00} --- Static SOK
}\label{Fig:SOK}
\end{figure}
\bigskip
\noindent\underline{\textbf{Important observation $\#1$:}} $aQ_B
\longrightarrow e(S_A,Q_B)$.

\subsection{Semi-Static and Ephemeral SOK Protocols}
\subsubsection{The Semi-Static SOK Protocol}

\begin{figure}[H]
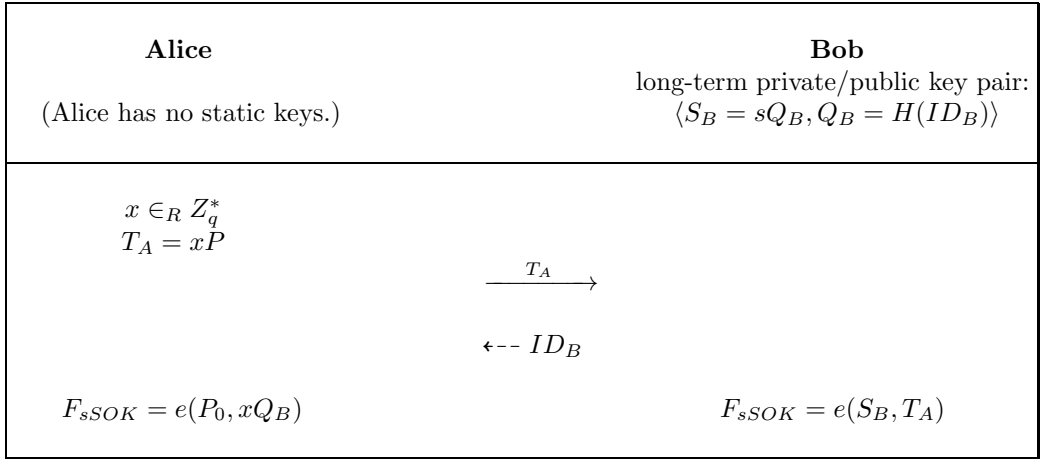

\center
\renewcommand{\arraystretch}{1}
\setlength\tabcolsep{0pt}

\begin{tabular}{|ccc|}
\hline
&&\\
\textbf{ Alice} & & \ \ \textbf{Bob}  \\
\ \              & &   \ \ long-term private/public key pair:\ \ \\
 \ \ \ \ (Alice has no static keys.)             & &   \ \ $\langle S_B = sQ_B, Q_B=H(ID_B)\rangle$\\
 &&\\
\hline
 &&\\
 $x \in_R Z_q^\ast$  & &  \ \ \\

$T_A = xP$   & &   \ \ \\

                 & \ \ \ \ \ \ \ \ \ \ \ \ \ \ $\xrightarrow{\ \ \ \ T_{A}\ \ \ \ }$ & \\
                 &&\\

                & \ \ \ \ \ \ \ \ \ \ \ \ \ \ \ \ $\dashleftarrow{ ID_{B}\ \ \ \ }$  & \\
                 & & \\

\ \ $F_{sSOK} = e(P_0, xQ_B)$ & &  \ \ $F_{sSOK} = e(S_B, T_A)$\ \  \\

&&\\
\hline

\end{tabular}
 \caption{The Semi-Static SOK Protocol}\label{Fig:SemiDynamic-SOK}
\end{figure}

It is well-known that the ElGamal encryption scheme \cite{El85} is
derived from the semi-static (or half-static, half-ephemeral)
Diffie-Hellman protocol \cite{MOV97}. Based on this seemingly
obvious relation, we find that the Boneh-Franklin ID-based
encryption (IBE) \cite{BF01,Sh84} is derived from the semi-static
SOK protocol (presented in Fig. \ref{Fig:SemiDynamic-SOK}). Note
that Paterson and Srinivasan \cite{PS07} also, independently,
noticed the relation. However, they do not give the term
``semi-static SOK protocol" explicitly (let alone the ephemeral SOK)
and only uses the static SOK protocol, \emph{i.e.} the SOK-NIKD
protocol. We stress that the explicit classification of the SOK
protocol, corresponding to the three version of the Diffie-Hellman
protocol, is essential for the main result of this paper.

In the rest of the paper, $P_0$ stands for the public key of the
private key generator (PKG), with $P_0=sP$ and $s$ being the master
private key of the PKG.

\subsubsection{The Ephemeral SOK Protocol}

\begin{figure}[H]
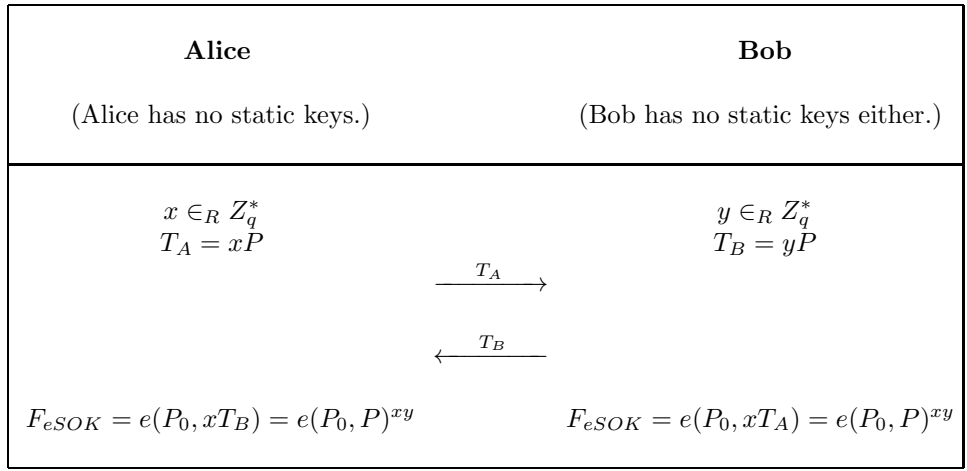

\center
\renewcommand{\arraystretch}{1}
\setlength\tabcolsep{0pt}

\begin{tabular}{|ccc|}
\hline
&&\\
\textbf{ Alice} & & \ \ \textbf{Bob}  \\
  & &   \ \ \\
\ \ (Alice has no static keys.)    & &   \ \ (Bob has no static keys either.)\ \ \\

 &&\\
\hline
 &&\\
 $x \in_R Z_q^\ast$  & &  \ \ $y \in_R Z_q^\ast$\\

$T_A = xP$   & &   \ \ $T_B = yP$\\

                 &  \ \ $\xrightarrow{\ \ \ \ T_{A}\ \ \ \ }$ & \\
                 &&\\

                &  \ \ $\xleftarrow{\ \ \ \ T_{B}\ \ \ \ }$  & \\
                 & & \\

\ \ $F_{eSOK} = e(P_0, xT_B) = e(P_0, P)^{xy}$ & &  \ \ $F_{eSOK} = e(P_0, xT_A)= e(P_0, P)^{xy}$\ \  \\

&&\\
\hline

\end{tabular}
 \caption{Ephemeral SOK Protocol}\label{Fig:Dynamic-SOK}
\end{figure}

The protocol is presented in Fig. \ref{Fig:Dynamic-SOK}.

\subsection{The UM and the RYY Protocols}
The RYY protocol \cite{RYY04} is build upon the UM protocol
\cite{AJM95,Go90}\footnote{Later, however, we will see that in the
exact ID-based version of the UM protocol, $xyP$ should be replaced
by $e(xsP,yp)$. This creates an escrowable RYY protocol.}. The two
session secrets of the two protocols are $K=F_{DH}||xyP$ and
$K=F_{SOK}||xyP$, respectively. A common weakness of them is that
they do not possess K-CI resilience \cite{BC05,WCB05}.

\begin{figure}[H]
\center
\renewcommand{\arraystretch}{1}
\setlength\tabcolsep{0pt}

\begin{tabular}{|ccc|}
\hline
&&\\
\textbf{ Alice} & & \ \ \textbf{Bob}  \\
\ \ long-term private/public key pair:    & &   \ \ long-term private/public key pair:\ \ \\
 $\langle a, Q_A = aP\rangle$   & &   \ \ $\langle b, Q_B = bP\rangle$\\
 &&\\
\hline
 &&\\
 $x \in_R Z_q^\ast$  & &  \ \ $y \in_R Z_q^\ast$\\

$T_A = xP$   & &   \ \ $T_B = yP$\\

                 &  \ \ $\xrightarrow{\ \ \ \ T_{A}\ \ \ \ }$ & \\
                 &&\\

                &  \ \ $\xleftarrow{\ \ \ \ T_{B}\ \ \ \ }$  & \\
                 & & \\

\ \ $F_{DH} = \mathbf{aQ_B}=abP$ & &  \ \ $F_{DH}  = bQ_A = abP$\ \  \\
\ \ $k = xT_B=xyP$ & &  \ \ $k= yT_A=xyP$\ \  \\
\ \ $sk = H_2(A||B||F_{DH}||k)$ & &  \ \ $sk = H_2(A||B||F_{DH}||k)$\ \  \\

&&\\
\hline

\end{tabular}
\caption{The UM Protocol \cite{AJM95}}\label{Fig:UMP}
\end{figure}


\begin{figure}[H]
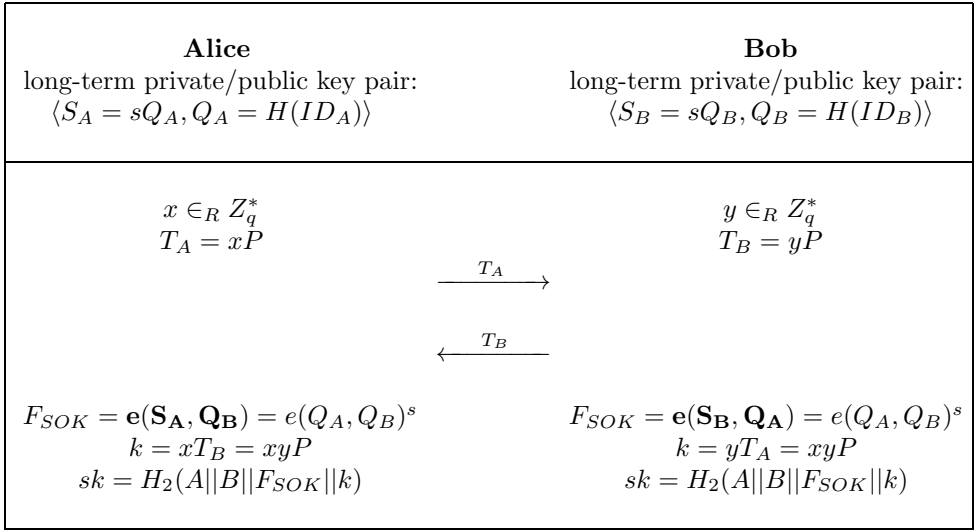

\center
\renewcommand{\arraystretch}{1}
\setlength\tabcolsep{0pt}

\begin{tabular}{|ccc|}
\hline
&&\\
\textbf{ Alice} & & \ \ \textbf{Bob}  \\
\ \ long-term private/public key pair:    & &   \ \ long-term private/public key pair:\ \ \\
 $\langle S_A = sQ_A, Q_A=H(ID_A)\rangle$   & &   \ \ $\langle S_B = sQ_B, Q_B=H(ID_B)\rangle$\\
 &&\\
\hline
 &&\\
 $x \in_R Z_q^\ast$  & &  \ \ $y \in_R Z_q^\ast$\\

$T_A = xP$   & &   \ \ $T_B = yP$\\

                 &  \ \ $\xrightarrow{\ \ \ \ T_{A}\ \ \ \ }$ & \\
                 &&\\

                &  \ \ $\xleftarrow{\ \ \ \ T_{B}\ \ \ \ }$  & \\
                 & & \\

\ \ $F_{SOK} = \mathbf{e(S_A, Q_B)}=e(Q_A, Q_B)^s$ & &  \ \ $F_{SOK} = \mathbf{e(S_B, Q_A)}=e(Q_A, Q_B)^s$\ \  \\
\ \ $k = xT_B=xyP$ & &  \ \ $k = yT_A=xyP$\ \  \\
\ \ $sk = H_2(A||B||F_{SOK}||k)$ & &  \ \ $sk = H_2(A||B||F_{SOK}||k)$\ \  \\

&&\\
\hline

\end{tabular}
\caption{The RYY Protocol \cite{RYY04}}\label{Fig:RYY}
\end{figure}

\subsection{The MTI/A0 and the Smart Protocols}

For those who are unfamiliar with the MTI protocol family, we refer
to \cite{MOV97,BM99,BJM97}. The same design idea that produces the
MTI/A0 and the Smart protocols was previously noticed, e.g. in
\cite{WCC06}, the authors used the term ``Encrypt--Decrypt method".
Concretely, the MTI/A0 protocol is based on the standard ElGamal
encryption, while Smart's protocol \cite{Sm02} is based on the
Boneh--Franklin IBE \cite{BF01}. However, the relations between the
computation of the two session secrets (c.f. the following
observation No. 2) has not yet been identified before. The two
session secrets of the two protocols are $K=aT_B + xQ_B$ and
$K=e(S_A, T_B)e(sP, xQ_B)$, respectively. A common weakness of the
two protocol is that they do not have perfect forward secrecy (PFS).

\begin{figure}[H]
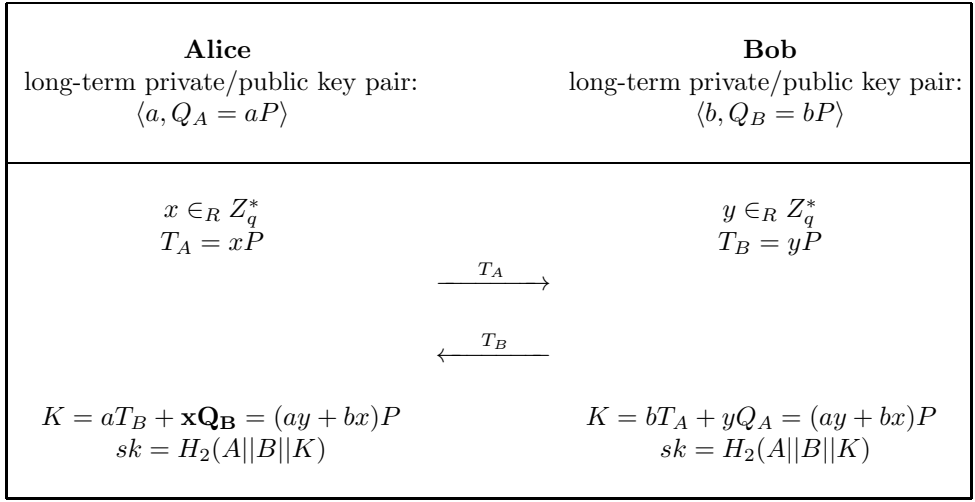

\center
\renewcommand{\arraystretch}{1}
\setlength\tabcolsep{0pt}

\begin{tabular}{|ccc|}
\hline
&&\\
\textbf{ Alice} & & \ \ \textbf{Bob}  \\
\ \ long-term private/public key pair:    & &   \ \ long-term private/public key pair:\ \ \\
 $\langle a, Q_A = aP\rangle$   & &   \ \ $\langle b, Q_B = bP\rangle$\\
 &&\\
\hline
 &&\\
 $x \in_R Z_q^\ast$  & &  \ \ $y \in_R Z_q^\ast$\\

$T_A = xP$   & &   \ \ $T_B = yP$\\

                 &  \ \ $\xrightarrow{\ \ \ \ T_{A}\ \ \ \ }$ & \\
                 &&\\

                &  \ \ $\xleftarrow{\ \ \ \ T_{B}\ \ \ \ }$  & \\
                 & & \\

\ \ $K = aT_B + \mathbf{xQ_B} = (ay+bx)P$ & &  \ \ $K = bT_A + yQ_A = (ay+bx)P$ \ \  \\
\ \ $sk = H_2(A||B||K)$ & &  \ \ $sk = H_2(A||B||K)$\ \  \\

&&\\
\hline

\end{tabular}
\caption{The MTI/A0 Protocol \cite{MTI86}}\label{Fig:A0}
\end{figure}


\begin{figure}[H]
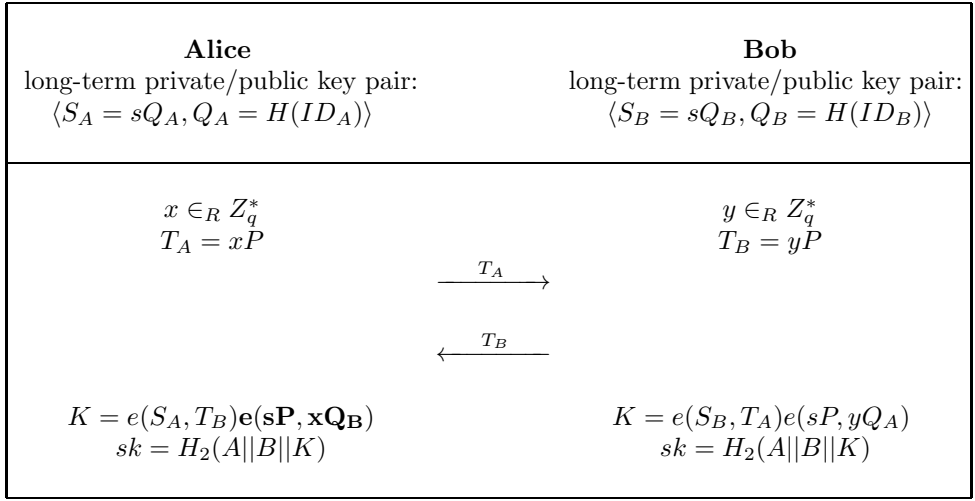

\center
\renewcommand{\arraystretch}{1}
\setlength\tabcolsep{0pt}

\begin{tabular}{|ccc|}
\hline
&&\\
\textbf{ Alice} & & \ \ \textbf{Bob}  \\
\ \ long-term private/public key pair:    & &   \ \ long-term private/public key pair:\ \ \\
 $\langle S_A = sQ_A, Q_A=H(ID_A)\rangle$   & &   \ \ $\langle S_B = sQ_B, Q_B=H(ID_B)\rangle$\\
 &&\\
\hline
 &&\\
 $x \in_R Z_q^\ast$  & &  \ \ $y \in_R Z_q^\ast$\\

$T_A = xP$   & &   \ \ $T_B = yP$\\

                 &  \ \ $\xrightarrow{\ \ \ \ T_{A}\ \ \ \ }$ & \\
                 &&\\

                &  \ \ $\xleftarrow{\ \ \ \ T_{B}\ \ \ \ }$  & \\
                 & & \\

\ \ $K = e(S_A,T_B)\mathbf{e(sP, xQ_B)}$ & &  \ \ $K = e(S_B,T_A)e(sP, yQ_A)$  \ \  \\
\ \ $sk = H_2(A||B||K)$ & &  \ \ $sk = H_2(A||B||K)$\ \  \\

&&\\
\hline

\end{tabular}
\caption{The Smart Protocol \cite{Sm02}}\label{Fig:Smart}
\end{figure}

From our first observation, $aT_B$ should be changed to $e(S_A,
T_B)$. Here we further notice that $xQ_B$ is changed to $e(sP,
xQ_B)$, with the help of the master public-key $P_0$ ($P_0=sP$)
\footnote{In \cite{WC07}, it was shown that under the SOK key
setting, IBE also exists if the master public-key of the PKG is set
to be $P_0=\mathbf{s^{-1}}P$. We stress that this is also true with
ID-based key agreement protocols, namely setting $P_0=s^{-1}P$ will
\emph{not} affect the correctness and security of the A0 type
ID-based protocols (e.g., Smart's, the SYL and our proposed
\ID-MQV), all that needed is to replace the protocol message
$T_A=xP$ with $T_A=xP_0$, and then adjust the computation of the
session secrets accordingly.}. Therefore, we get our second
observation. Here $Q_i$ ($i=\{1,2\}$) are any publicly computable
elements in group $\mathbb{G}_1$, such as $Q_A+Q_B$, $Q_A+T_B$, with
$Q_A$, $Q_B$ being public keys and $T_B$ being the protocol message
sent out by Bob.

\bigskip
\noindent\underline{\textbf{Important observation $\#2$:}}
$aQ_1+xQ_2 \longrightarrow e(S_A, Q_1)e(P_0, xQ_2)$.\\

We summarize the above two observations with the following two
substitution rules in Table \ref{Tab:SOK}.

\begin{table}[H]
  \centering
  \caption{Substitution Rules for the SOK Key Construction }\label{Tab:SOK}
  \begin{tabular}{|l|lll|c|}
    \hline
   & Auth. DH                    &  &  ID-Based Protocols \\
    &&&\\
   & Static Private-key: a       & & Static Private-key: $S_A=sQ_A$ \\
 Notations   & Static Public-key: $Q_A=aP$ & & Static Public-key: $Q_A=H(ID_A)$ \\
    & Ephemeral Private-key: x       & & Ephemeral Private-key: $x$ \\
    & Publicly-computable group element: & & Publicly-computable group element: \ \ \ \\
    & \ \  $Q,Q_1,Q_2$ & &\ \  $Q,Q_1,Q_2$ \\
   \hline
    &&&\\
  Two Rules  & Rule 1.  $ K=aQ$                    & \ \ $\Leftrightarrow$ \ \ & $K=e(S_A, Q)$ \\
             & Rule 2. $ K=aQ_1+xQ_2$             & \ \  $\Leftrightarrow$ \ \ &$K=e(S_A,Q_1)e(P_0,xQ_2)$   \\
&&&\\
    \hline
  \end{tabular}

\end{table}

\section{Relations between Pairs of Existing Protocols}

Applying the above two important substitution rules, we discover
some unpublished relations between some pairs of existing protocols.

\subsection{The MTI/A1 and the Chen--Kudla Protocols}
The Chen--Kudla protocol \cite{CK02} can be obtained by directly
applying the above two substitution rules. In MTI/A1, the session
secret is $K= aT_B+axQ_B$. Therefore in its ID-based counterpart,
the session secret is $K=e(S_A, T_B)e(S_A, xQ_B)=e(S_A, T_B+xQ_B)$.
This is exactly the Chen--Kudla \cite{CK02} protocol!

\begin{figure}[H]
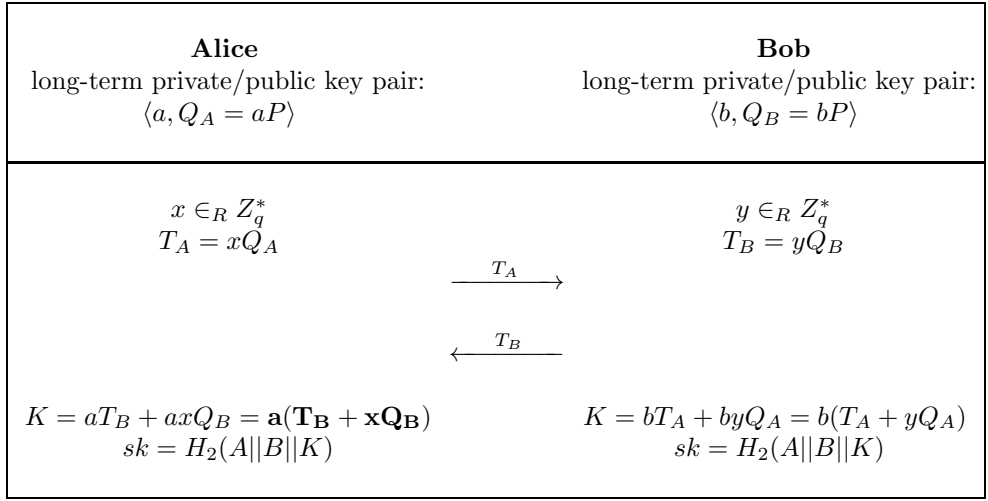

\center
\renewcommand{\arraystretch}{1}
\setlength\tabcolsep{0pt}

\begin{tabular}{|ccc|}
\hline
&&\\
\textbf{ Alice} & & \ \ \textbf{Bob}  \\
\ \ long-term private/public key pair:    & &   \ \ long-term private/public key pair:\ \ \\
 $\langle a, Q_A = aP\rangle$   & &   \ \ $\langle b, Q_B = bP\rangle$\\
 &&\\
\hline
 &&\\
 $x \in_R Z_q^\ast$  & &  \ \ $y \in_R Z_q^\ast$\\

$T_A = xQ_A$   & &   \ \ $T_B = yQ_B$\\

                 &  \ \ $\xrightarrow{\ \ \ \ T_{A}\ \ \ \ }$ & \\
                 &&\\

                &  \ \ $\xleftarrow{\ \ \ \ T_{B}\ \ \ \ }$  & \\
                 & & \\

\ \ $K = aT_B + axQ_B = \mathbf{a(T_B + xQ_B)}$ & &  \ \ $K = bT_A + byQ_A = b(T_A + yQ_A)$ \ \  \\
\ \ $sk = H_2(A||B||K)$ & &  \ \ $sk = H_2(A||B||K)$\ \  \\

&&\\
\hline

\end{tabular}
\caption{The MTI/A1 Protocol \cite{MTI86}}\label{Fig:A1}
\end{figure}


\begin{figure}[H]
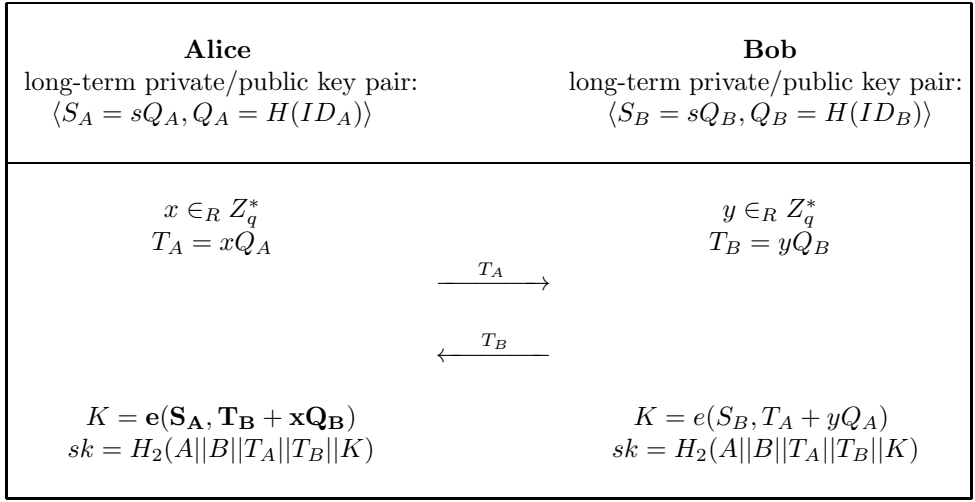

\center
\renewcommand{\arraystretch}{1}
\setlength\tabcolsep{0pt}

\begin{tabular}{|ccc|}
\hline
&&\\
\textbf{ Alice} & & \ \ \textbf{Bob}  \\
\ \ long-term private/public key pair:    & &   \ \ long-term private/public key pair:\ \ \\
 $\langle S_A = sQ_A, Q_A=H(ID_A)\rangle$   & &   \ \ $\langle S_B = sQ_B, Q_B=H(ID_B)\rangle$\\
 &&\\
\hline
 &&\\
 $x \in_R Z_q^\ast$  & &  \ \ $y \in_R Z_q^\ast$\\

$T_A = xQ_A$   & &   \ \ $T_B = yQ_B$\\

                 &  \ \ $\xrightarrow{\ \ \ \ T_{A}\ \ \ \ }$ & \\
                 &&\\

                &  \ \ $\xleftarrow{\ \ \ \ T_{B}\ \ \ \ }$  & \\
                 & & \\

\ \ $K = \mathbf{e(S_A,T_B + xQ_B)}$ & &  \ \ $K = e(S_B,T_A + yQ_A)$  \ \  \\
\ \ $sk = H_2(A||B||T_A||T_B||K)$ & &  \ \ $sk = H_2(A||B||T_A||T_B||K)$\ \  \\

&&\\
\hline

\end{tabular}
\caption{The Chen--Kudla Protocol \cite{CK02}}\label{Fig:Chen-Kudla}
\end{figure}

\subsection{The MQV-1 and Wang's Protocols}
Wang's protocol \cite{Wangyongge05} can be obtained from the
so-called MQV-1 protocol by directly applying the above two rules.

We first review the famous MQV \cite{LMQSV03} protocol. Note that
the HMQV protocol \cite{Kra05} is a hashed variant of the MQV
protocol.

\begin{figure}[H]
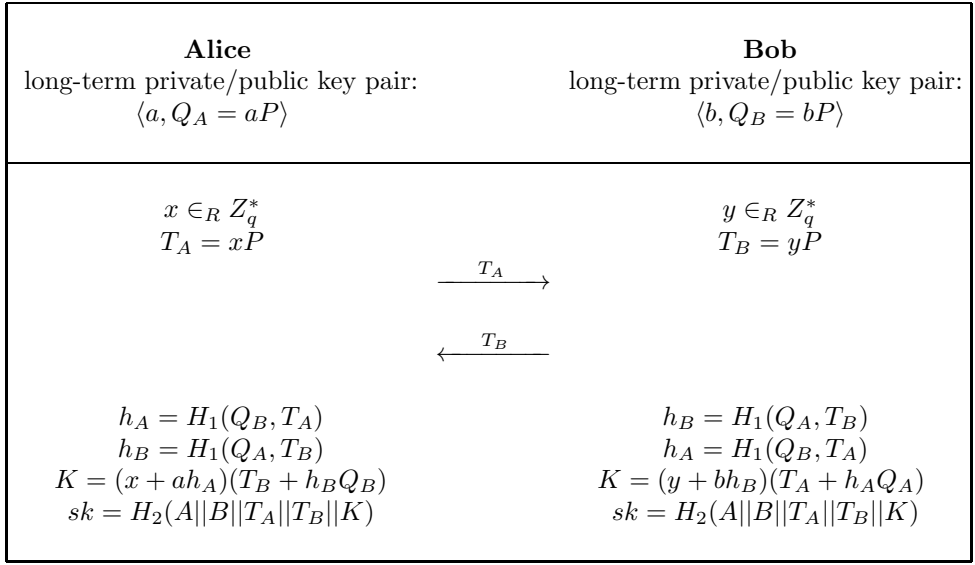

\center
\renewcommand{\arraystretch}{1}
\setlength\tabcolsep{0pt}

\begin{tabular}{|ccc|}
\hline
&&\\
\textbf{ Alice} & & \ \ \textbf{Bob}  \\
\ \ long-term private/public key pair:    & &   \ \ long-term private/public key pair:\ \ \\
 $\langle a, Q_A = aP\rangle$   & &   \ \ $\langle b, Q_B = bP\rangle$\\
 &&\\
\hline
 &&\\
 $x \in_R Z_q^\ast$  & &  \ \ $y \in_R Z_q^\ast$\\

$T_A = xP$   & &   \ \ $T_B = yP$\\

                 &  \ \ $\xrightarrow{\ \ \ \ T_{A}\ \ \ \ }$ & \\
                 &&\\

                &  \ \ $\xleftarrow{\ \ \ \ T_{B}\ \ \ \ }$  & \\
                 & & \\
\ \ $h_A = H_1(Q_B, T_A)$ & &  \ \ $h_B = H_1(Q_A, T_B)$\ \  \\
\ \ $h_B = H_1(Q_A, T_B)$ & &  \ \ $h_A = H_1(Q_B, T_A)$\ \  \\
\ \ $K = (x+ah_A)(T_B+h_B Q_B)$ & &  \ \ $K = (y+bh_B)(T_A+h_A Q_A)$ \ \  \\
\ \ $sk = H_2(A||B||T_A||T_B||K)$ & &  \ \ $sk = H_2(A||B||T_A||T_B||K)$\ \  \\

&&\\
\hline

\end{tabular}
\caption{The (H)MQV Protocol \cite{LMQSV03,Kra05}}\label{Fig:MQV}
\end{figure}


The MQV-1 protocol can be obtained by simply changing the protocol
message $T_A=xP$ to be $T_A=xQ_A$, and then adjust the protocol
accordingly. The session secret of the MQV-1 protocol is $K=
(x+h_A)a(T_B+h_BQ_B)$. Therefore in its ID-based counterpart, the
session secret is $K=e((x+h_A)S_A, T_B+h_BQ_B)$, this is exactly the
Chow--Choo protocol \cite{CC07}
--- a hashed variant of Wang's protocol \cite{Wangyongge05}.

\section{Obtaining the Real ID-Based MQV Protocol}
\subsection{Our \ID-MQV Protocol}

The session secret in (H)MQV is as follows: $$K =
(x+h_Aa)(T_B+h_BQ_B)=x(T_B+h_BQ_B)+h_Aa(T_B+h_BQ_B).$$ We let
$Q_1=T_B+h_BQ_B$ and $Q_2=h_A(T_B+h_BQ_B)=h_AQ_1$, then
$$K=xQ_1+aQ_2,$$
Applying Rule \#2, we obtain the ID-based version of this protocol
--- \ID-MQV, its session secret $K$ is as follows:

$$K =
e(P_0,xQ_1)e(S_A,Q_2)=e(xP_0,Q_1)e(h_AS_A,Q_1)=
e(xP_0+h_AS_A,Q_1),$$ recall that $Q_1=T_B+h_BQ_B$, thus we have

$$K= e(xP_0+h_AS_A,T_B+h_BQ_B).$$

\begin{figure}[H]
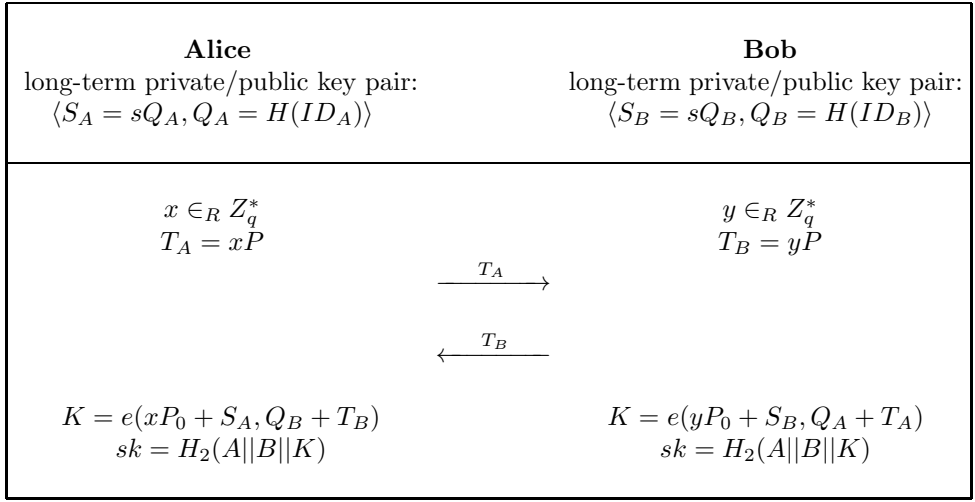

\center
\renewcommand{\arraystretch}{1}
\setlength\tabcolsep{0pt}

\begin{tabular}{|ccc|}
\hline
&&\\
\textbf{ Alice} & & \ \ \textbf{Bob}  \\
\ \ long-term private/public key pair:    & &   \ \ long-term private/public key pair:\ \ \\
 $\langle S_A = sQ_A, Q_A=H(ID_A)\rangle$   & &   \ \ $\langle S_B = sQ_B, Q_B=H(ID_B)\rangle$\\
 &&\\
\hline
 &&\\
 $x \in_R Z_q^\ast$  & &  \ \ $y \in_R Z_q^\ast$\\

$T_A = xP$   & &   \ \ $T_B = yP$\\

                 &  \ \ $\xrightarrow{\ \ \ \ T_{A}\ \ \ \ }$ & \\
                 &&\\

                &  \ \ $\xleftarrow{\ \ \ \ T_{B}\ \ \ \ }$  & \\
                 & & \\

 \ \ $K = e(x P_{0} + S_A, Q_B + T_B)$ & &  \ \ $K = e(y P_{0} + S_B, Q_A + T_A)$\ \  \\
 \ \ $sk = H_2(A||B||K)$ & &  \ \ $sk = H_2(A||B||K)$\ \  \\

&&\\
\hline

\end{tabular}
\caption{The Shim Protocol \cite{Sh03}}\label{Fig:Shim}
\end{figure}

If we wipe off $h_A$ and $h_B$, then the above \ID-MQV protocol
degenerate into the Shim  protocol \cite{Sh03} which is given in Fig
\ref{Fig:Shim}. However, the Shim protocols is totally broken by Sun
and Hsie \cite{SH03}. In 2005, Yuan and Li \cite{YL05} repaired the
Shim protocol using a very simple idea, namely just adding an
ephemeral Diffie-Hellman value. The improved protocol is called the
Shim-Yuan-Li (SYL) protocol (see Fig. \ref{Fig:SYL}) and was proven
to be secure by Chen \emph{et al.} \cite{CCS06}. In Fig.
\ref{Fig:nonid-SYL}, we present the non-ID-based version of the SYL
protocol --- \nID-SYL.

\subsection{Remarks on the \ID-MQV Protocol}

\begin{figure}[H]
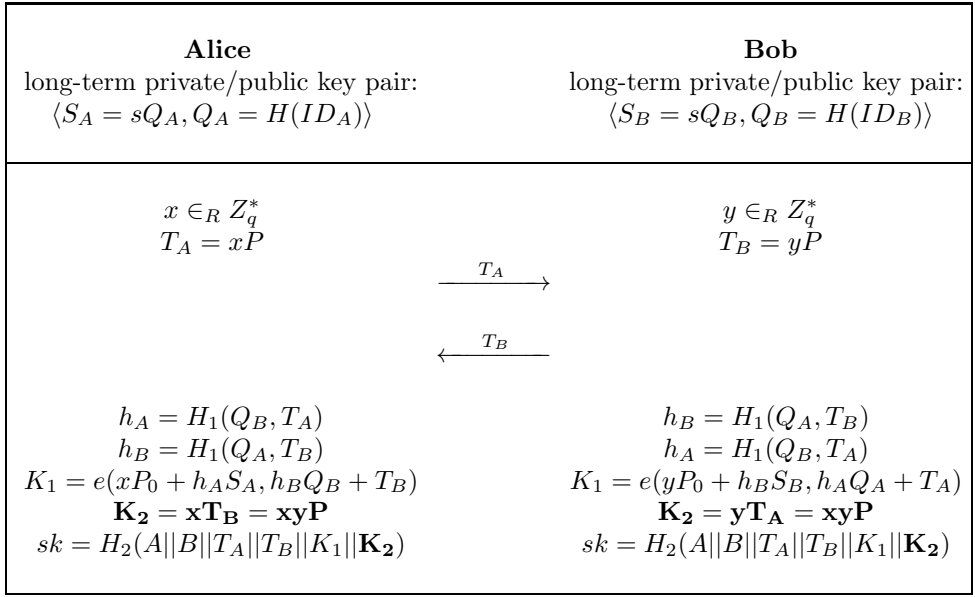

\center
\renewcommand{\arraystretch}{1}
\setlength\tabcolsep{0pt}

\begin{tabular}{|ccc|}
\hline
&&\\
\textbf{ Alice} & & \ \ \textbf{Bob}  \\
\ \ long-term private/public key pair:    & &   \ \ long-term private/public key pair:\ \ \\
 $\langle S_A = sQ_A, Q_A=H(ID_A)\rangle$   & &   \ \ $\langle S_B = sQ_B, Q_B=H(ID_B)\rangle$\\
 &&\\
\hline
 &&\\
 $x \in_R Z_q^\ast$  & &  \ \ $y \in_R Z_q^\ast$\\

$T_A = xP$   & &   \ \ $T_B = yP$\\

                 &  \ \ $\xrightarrow{\ \ \ \ T_{A}\ \ \ \ }$ & \\
                 &&\\

                &  \ \ $\xleftarrow{\ \ \ \ T_{B}\ \ \ \ }$  & \\
                 & & \\

\ \ $h_A = H_1(Q_B, T_A)$ & &  \ \ $h_B = H_1(Q_A, T_B)$\ \  \\
\ \ $h_B = H_1(Q_A, T_B)$ & &  \ \ $h_A = H_1(Q_B, T_A)$\ \  \\
  \ \ $K_1 = e(x P_{0} + h_A S_A, h_B Q_B + T_B)$ & &  \ \ $K_1 = e(y P_{0} + h_BS_B, h_AQ_A + T_A)$\ \  \\
 \ \ $\mathbf{K_2 = xT_B=xyP}$ & &  \ \ $\mathbf{K_2 = yT_A=xyP}$\ \  \\
 \ \ $sk = H_2(A||B||T_A||T_B||K_1||\mathbf{K_2})$ & &  \ \ $sk = H_2(A||B||T_A||T_B||K_1||\mathbf{K_2})$\ \  \\

&&\\
\hline
\end{tabular}\\

\caption{Escrowless \ID-MQV: ID-Based (H)MQV Protocol with
PKG-FS}\label{Fig:PKGFS-ID-MQV}
\end{figure}

Our \ID-MQV protocol has remarkable superiorities over all the
existing ID-based key agreement protocols (from pairings).

\begin{enumerate}
  \item From the format of the protocol messages, we argue that our \ID-MQV is the real ID-based
  version of the famous (H)MQV protocol. As mentioned above, the
  Chow--Choo and Wang protocols are ID-based version of the
  so-called (H)MQV-1 protocols, which have different protocol
  messages.\smallskip\smallskip
  \item Separating  \emph{perfect forward secrecy} (PFS) from \emph{PKG forward
  secrecy} (PKG-FS). Note that PKG-FS also means escrowless. We argue that in
  some applications (as also pointed out by McCullagh and Barreto \cite{MB05a}) key escrow is a
  requirement or even, a must. However, if we
  remove $K_1=abP$ from the SYL protocol \cite{YL05} to open escrow, then it
  become totally insecure (which is exactly Shim's protocol \cite{Sh03}), let
  alone PFS. Our new protocol can be securely used in escrowed model
  (\emph{i.e.}, w/o $xyP$), providing PFS. When $xyP$ is added, the
  protocol becomes escrowless (and achieves PKG-FS, see Fig. \ref{Fig:PKGFS-ID-MQV}). In a word, $xyP$
  separates clearly PFS from PKG-FS, and our new protocol (\ID-MQV) can be used
  with or without escrow.\smallskip\smallskip
  \item Compared with Wang's protocol \cite{Wangyongge05} (and the Chow-Choo protocol \cite{CC07}), our protocol does not
  need extra message exchange to close escrow, while the latter
  requires a party to send out an extra point. At the same time,
  brings extra computation for the party.\smallskip\smallskip
  \item The new protocol can be further strengthened to achieve stronger
  security, \emph{i.e.}, to be secure in the extended Canetti--Krawczyk (eCK)
model which allows \emph{ephemeral }secret key reveal.
  (Using the same idea from \cite{LLM06}.)
\end{enumerate}

\section{Beyond the SOK ID-Based Key Construction}

Now we look at the SK key setting. For details on the key setting,
please refer to \cite{SK03} and \cite{MB05a,Xie05a}. Here we note
that the master private and public key pair of the PKG is $\langle
s, P_0=sP\rangle$. $u$ is part of a user's static public key and for
Alice $u_A=H'(ID_A)\in \mathbb{Z}_q^\ast$.

We discover that the key transport protocol behind the SK-IBE
\cite{SK03} is simply the ID-based version of the Hughes protocol
\cite{Hug94}. This is mainly because the static private key of the
receivers in the two protocols are both inversion-based. The
substitution rules are listed in Table \ref{Tab:SK}.

\begin{table}[H]
  \centering
  \caption{Substitution Rules for the SK Key Construction }\label{Tab:SK}
  \begin{tabular}{|l|lcl|}
    \hline
   & Auth. DH                    &  &  ID-Based Protocols \\
    &&&\\
   & Static key pair:        & & Static key pair: $\langle S_A=s+u_A)^{-1}Q_P,$\ \ \\
 Notations   &\ \  $\langle a, Q_A=aP\rangle$ & &\ \  $Q_A=P_0+u_AP=(s+u_A)P\rangle$ \\
    & Ephemeral Private-key: x       & & Ephemeral Private-key: $x$ \\
    & Publicly-computable element: $Q$ & & Publicly-computable element: $Q$ \\
   \hline
    &&&\\
              & Rule 1.  $ K=a^{-1}Q$ &\ \ \ \ $\Leftrightarrow$\ \ \ \ & $K=e(S_A, Q)$ \\
    Two Rules  &  Rule 2. $K=xP$,     &\ \ \ \ $\Leftrightarrow$\ \ \ \ & $K=e(P,P)^x$ \\
              &     &  &  \\
        \hline
  \end{tabular}

\end{table}

Using the above rules, we can establish the relations between the MB
protocols \cite{MB05a,MB05b} and the MTI/C0 and MTI/B0 \cite{MTI86}
protocols (c.f. Table \ref{Tab:2}), the details are omitted here.
Next, based on the enhanced MTI/C0 protocol (\emph{i.e.} the ECKE-1N
protocol), we propose a highly efficient ID-based protocol --- \eMB.

\subsection{Review of the ECKE-1N Protocol}

\begin{figure}[H]
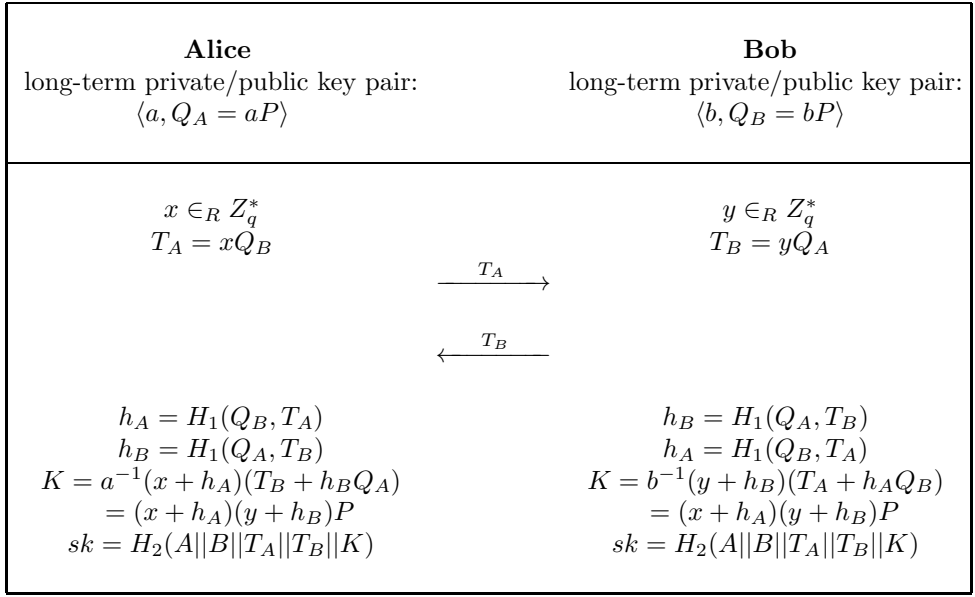

\center
\renewcommand{\arraystretch}{1}
\setlength\tabcolsep{0pt}

\begin{tabular}{|ccc|}
\hline
&&\\
\textbf{ Alice} & & \ \ \textbf{Bob}  \\
\ \ long-term private/public key pair:    & &   \ \ long-term private/public key pair:\ \ \\
 $\langle a, Q_A = aP\rangle$   & &   \ \ $\langle b, Q_B = bP\rangle$\\
 &&\\
\hline
 &&\\
 $x \in_R Z_q^\ast$  & &  \ \ $y \in_R Z_q^\ast$\\

$T_A = xQ_B$   & &   \ \ $T_B = yQ_A$\\

                 &  \ \ $\xrightarrow{\ \ \ \ T_{A}\ \ \ \ }$ & \\
                 &&\\

                &  \ \ $\xleftarrow{\ \ \ \ T_{B}\ \ \ \ }$  & \\
                 & & \\
\ \ $h_A = H_1(Q_B, T_A)$ & &  \ \ $h_B = H_1(Q_A, T_B)$\ \  \\
\ \ $h_B = H_1(Q_A, T_B)$ & &  \ \ $h_A = H_1(Q_B, T_A)$\ \  \\
\ \ $K = a^{-1}(x+h_A)(T_B+h_BQ_A)$ & &  \ \ $K = b^{-1}(y+h_B)(T_A+h_AQ_B)$\ \  \\
\ \ $\ \  = (x+h_A) (y+h_B)P$ & &  \ \ $\ \  = (x+h_A) (y+h_B)P$\ \  \\
 \ \ $sk = H_2(A||B||T_A||T_B||K)$ & &  \ \ $sk = H_2(A||B||T_A||T_B||K)$\ \  \\

&&\\
\hline

\end{tabular}
\caption{The Enhanced MTI/C0 Protocol --- ECKE-1N
}\label{Fig:Enhanced-C0}
\end{figure}

This protocol was initially designed using the ideas from MQV. It
was later included in a Letter appeared in IEEE Communications
Letters entitled ``Cryptanalysis and Improvement of an Elliptic
Curve Diffie-Hellman Protocol" \cite{WCSW08}. (Also available at
IACR ePrint, report 2007/026.) The protocol is give in Fig.
\ref{Fig:Enhanced-C0}.

\subsection{The \eMB \ Protocol}

Applying the substitution rules from Table \ref{Tab:SK}, we converse
our ECKE-1N into an ID-based authenticated key agreement protocol
which is the enhanced version of the McCullagh-Barreto protocol
\cite{MB05a,MB05b} --- \eMB.

\begin{figure}[H]
\center
\renewcommand{\arraystretch}{1}
\setlength\tabcolsep{0pt}

\begin{tabular}{|ccc|}
\hline
&&\\
\textbf{ Alice} & & \ \ \textbf{Bob}  \\
\ \ long-term private/public key pair:    & &   \ \ long-term private/public key pair:\ \ \\
\ \ \ \  $\langle S_A = (s+u_A)^{-1}P$ ,   & &   \ \  $\langle S_B = (s+u_B)^{-1}P$,  \ \ \ \ \\
\ \ \ \  $Q_A=P_0+u_AP=(s+u_A)P\rangle$          & &   \ \ $Q_B=P_0+u_BP=(s+u_B)P\rangle$ \ \ \ \ \\
 &&\\
\hline
 &&\\
 $x \in_R Z_q^\ast$  & &  \ \ $y \in_R Z_q^\ast$\\
\ \ \ \  $Q_B=P_0+u_BP=(s+u_B)P$          & &   \ \ $Q_A=P_0+u_AP=(s+u_A)P$ \ \ \ \ \\
$T_A = xQ_B$   & &   \ \ $T_B = yQ_A$\\
 &&\\
                 &  \ \ $\xrightarrow{\ \ \ \ T_{A}\ \ \ \ }$ & \\
                 &&\\

                &  \ \ $\xleftarrow{\ \ \ \ T_{B}\ \ \ \ }$  & \\
                 & & \\

\ \ $h_A = H(Q_B, T_A)$ & &  \ \ $h_B = H(Q_A, T_B)$\ \  \\
\ \ $h_B = H(Q_A, T_B)$ & &  \ \ $h_A = H(Q_B, T_A)$\ \  \\
 \ \ $\ \ \ \ \ \ K = e((x + h_A) S_A, T_B + h_B Q_A)$ & &  \ \ $\ \ \ \ \ \ K = e((y + h_B) S_B, T_A + h_A Q_B)$  \ \  \\
 \ \ $= e(P,P)^{(x + h_A)(y + h_B)}$ & &  \ \ $= e(P,P)^{(x + h_A)(y + h_B)}$  \ \  \\
 \ \ $sk = H_2(A||B||T_A||T_B||K)$ & &  \ \ $sk = H_2(A||B||T_A||T_B||K)$\ \  \\

&&\\
\hline
\end{tabular}\\

\caption{The \textsf{eMB} Protocol}\label{Fig:Enhanced-MB}
\end{figure}

We remark that the substitution rules in the SK ID-based key setting
can also be applied to the SK variants, e.g. Gentry's key setting
\cite{Gen06} and the second Boneh-Boyen ($BB_2$) scheme \cite{BB04}.

\section*{Acknowledgements}
The author wishes to thank Liqun Chen, Zhaohui Cheng, Kenny Paterson
and Yongge Wang for comments on the first version of this paper.

\begin{appendix}

\section{Obtaining an Authenticated DH Protocol from the SYL Protocol}
\label{Appendix:A}

\begin{figure}[H]
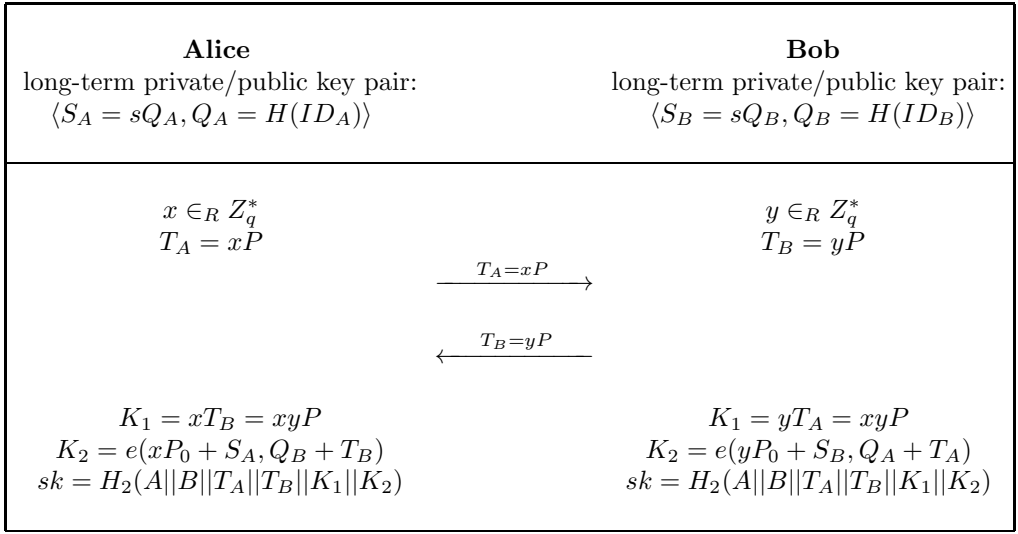

\center
\renewcommand{\arraystretch}{1}
\setlength\tabcolsep{0pt}

\begin{tabular}{|ccc|}
\hline
&&\\
\textbf{ Alice} & & \ \ \textbf{Bob}  \\
\ \ long-term private/public key pair:    & &   \ \ long-term private/public key pair:\ \ \\
 $\langle S_A = sQ_A, Q_A=H(ID_A)\rangle$   & &   \ \ $\langle S_B = sQ_B, Q_B=H(ID_B)\rangle$\\
 &&\\
\hline
 &&\\
 $x \in_R Z_q^\ast$  & &  \ \ $y \in_R Z_q^\ast$\\

$T_A = xP$   & &   \ \ $T_B = yP$\\

                 &  \ \ $\xrightarrow{\ \ \ \ T_{A}=xP\ \ \ \ }$ & \\
                 &&\\

                &  \ \ $\xleftarrow{\ \ \ \ T_{B}=yP\ \ \ \ }$  & \\
                 & & \\

\ \ $K_1 = xT_B=xyP$ & &  \ \ $K_1  = yT_A = xyP$\ \  \\
 \ \ $K_2 = e(x P_{0} + S_A, Q_B + T_B)$ & &  \ \ $K_2 = e(y P_{0} + S_B, Q_A + T_A)$\ \  \\
 \ \ $sk = H_2(A||B||T_A||T_B||K_1||K_2)$ & &  \ \ $sk = H_2(A||B||T_A||T_B||K_1||K_2)$\ \  \\

&&\\
\hline

\end{tabular}
\caption{The SYL Protocol \cite{YL05}}\label{Fig:SYL}
\end{figure}

\begin{figure}[H]
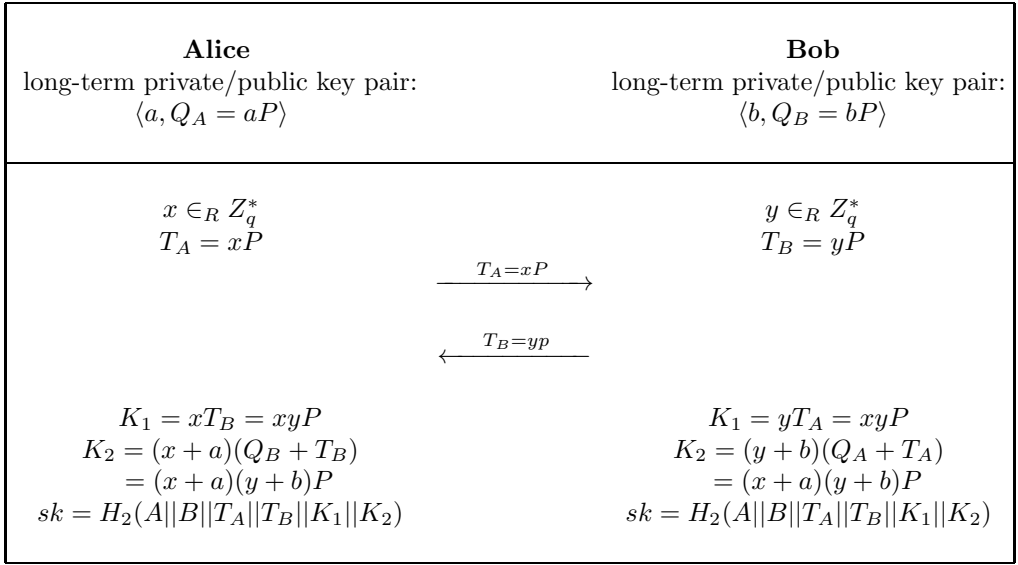

\center
\renewcommand{\arraystretch}{1}
\setlength\tabcolsep{0pt}

\begin{tabular}{|ccc|}
\hline
&&\\
\textbf{ Alice} & & \ \ \textbf{Bob}  \\
\ \ long-term private/public key pair:    & &   \ \ long-term private/public key pair:\ \ \\
 $\langle a, Q_A = aP\rangle$   & &   \ \ $\langle b, Q_B = bP\rangle$\\
 &&\\
\hline
 &&\\
 $x \in_R Z_q^\ast$  & &  \ \ $y \in_R Z_q^\ast$\\

$T_A = xP$   & &   \ \ $T_B = yP$\\

                 &  \ \ $\xrightarrow{\ \ \ \ T_{A}=xP\ \ \ \ }$ & \\
                 &&\\

                &  \ \ $\xleftarrow{\ \ \ \ T_{B} = yp\ \ \ \ }$  & \\
                 & & \\
\ \ $K_1 = xT_B=xyP$ & &  \ \ $K_1  = yT_A = xyP$\ \  \\
\ \ $K_2 = (x+a) (Q_B + T_B)$ & &  \ \ $K_2 = (y+b) (Q_A + T_A)$\ \  \\
\ \ $\ \  = (x+a) (y+b)P$ & &  \ \ $\ \  = (x+a) (y+b)P$\ \  \\
 \ \ $sk = H_2(A||B||T_A||T_B||K_1||K_2)$ & &  \ \ $sk = H_2(A||B||T_A||T_B||K_1||K_2)$\ \  \\

&&\\
\hline
\end{tabular}\\

\caption{\nID-SYL: A New Authenticated Diffie-Hellman
Protocol}\label{Fig:nonid-SYL}
\end{figure}

The two protocols are presented in Fig. \ref{Fig:SYL} and
\ref{Fig:nonid-SYL}, respectively.

\section{Enhanced MTI/C1 Protocol}
This protocol can be easily derived from our enhanced MTI/C0
protocol (\emph{i.e.} the ECKE-1N protocol) using the idea from
\cite{MTI86}.


\begin{figure}[H]
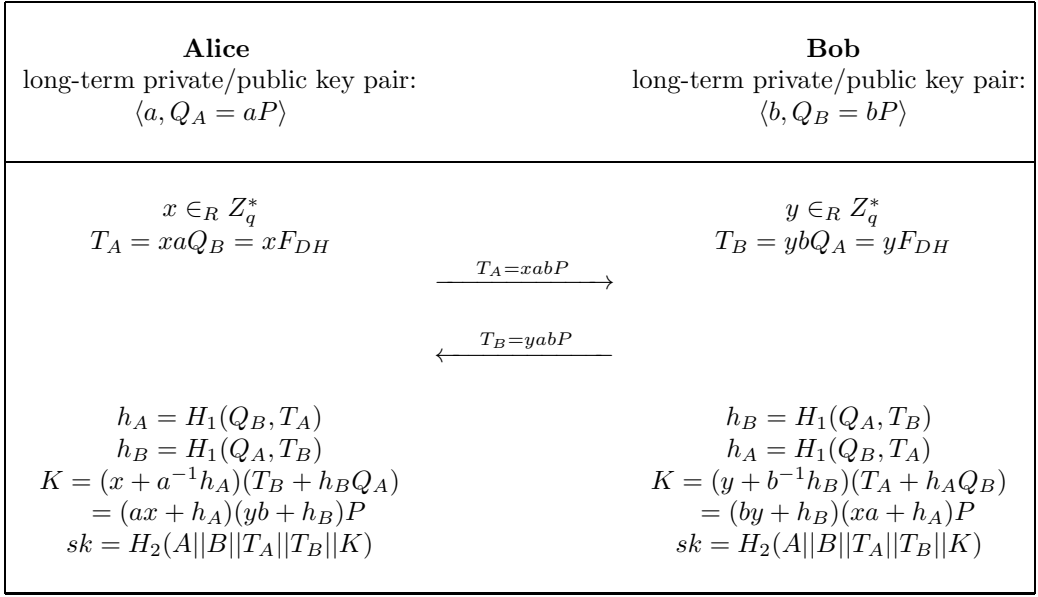

\center
\renewcommand{\arraystretch}{1}
\setlength\tabcolsep{0pt}

\begin{tabular}{|ccc|}
\hline
&&\\
\textbf{ Alice} & & \ \ \textbf{Bob}  \\
\ \ long-term private/public key pair:    & &   \ \ long-term private/public key pair:\ \ \\
 $\langle a, Q_A = aP\rangle$   & &   \ \ $\langle b, Q_B = bP\rangle$\\
 &&\\
\hline
 &&\\
 $x \in_R Z_q^\ast$  & &  \ \ $y \in_R Z_q^\ast$\\

$T_A =xaQ_B = xF_{DH}$   & &   \ \ $T_B =ybQ_A = yF_{DH}$\\

                 &  \ \ $\xrightarrow{\ \ \ \ T_{A}=xabP\ \ \ \ }$ & \\
                 &&\\

                &  \ \ $\xleftarrow{\ \ \ \ T_{B}=yabP\ \ \ \ }$  & \\
                 & & \\
\ \ $h_A = H_1(Q_B, T_A)$ & &  \ \ $h_B = H_1(Q_A, T_B)$\ \  \\
\ \ $h_B = H_1(Q_A, T_B)$ & &  \ \ $h_A = H_1(Q_B, T_A)$\ \  \\
\ \ $K = (x+a^{-1}h_A)(T_B+h_BQ_A)$ & &  \ \ $K = (y+b^{-1}h_B)(T_A+h_AQ_B)$\ \  \\
\ \ $\ \  = (ax+h_A) (yb+h_B)P$ & &  \ \ $\ \  = (by+h_B) (xa+h_A)P$\ \  \\
 \ \ $sk = H_2(A||B||T_A||T_B||K)$ & &  \ \ $sk = H_2(A||B||T_A||T_B||K)$\ \  \\

&&\\
\hline

\end{tabular}
\caption{The Enhanced MTI/C1 Protocol}\label{Fig:Enhanced-C1}
\end{figure}

\end{appendix}
\end{document}